\documentclass{JAC2003}

\usepackage{graphicx}

\begin{document}

\title{QUANTITATIVE DUALITY AND QUANTUM ERASURE FOR NEUTRAL KAONS
}

\author{A. Bramon, Grup de F{\'\i}sica Te\`orica,
Universitat Aut\'onoma de Barcelona, \\
E--08193 Bellaterra, Spain\\
G. Garbarino, Departament d'Estructura i Constituents de la
Mat\`{e}ria,
Universitat de Barcelona, \\ E--08028 Barcelona, Spain\\
B.C. Hiesmayr, Institute for Experimental Physics, University of Vienna, \\
A--1090 Vienna, Austria}

\maketitle

\begin{abstract}
The aim of this contribution is to illustrate two basic aspects of
quantum mechanics applied to the neutral kaon system. We first
describe a recent quantitative formulation of Bohr's
complementarity principle for free--space evolution of single
kaons and entangled kaon pairs. We then show that the neutral kaon
system is also suitable for an optimal demonstration of the
``quantum eraser'', including its operation in the ``delayed
choice'' mode. In our discussions, strangeness oscillations play
the role of the traditional interference pattern linked to
wave--like behaviour. The role of the two interferometric paths
taken by particle--like objects is played by the differently
propagating $K_S$ and $K_L$ components. Their distinct decay
widths provide a quantum ``mark'' which can be erased by
appropriate strangeness measurements.
\end{abstract}

\section{I.\, Introduction}

As it is well known, fundamental
aspects and principles of quantum mechanics can be conveniently illustrated
by the basic double--slit experiment \cite{feynman}.
In order to do this, one usually explores the
consequences of the complementarity relation existing
between the observation of interference fringes, conventionally
associated to wave--like behaviour, and
the acquisition of ``which way''
or {\it Welcher Weg} information, similarly associated to particle--like behaviour.

In a double--slit--like experiment,
interference patterns are observed if and only if
it is impossible to know, \emph{even in principle}, which path the particle took.
Interference disappears if there is a way to know ---e.g., through a
\emph{quantum marking} procedure--- which path
the particle took; whether or not the outcome of the corresponding ``which way''
observation is actually read out, it is completely irrelevant:
information is there and interference is in any way
lost. But, if that ``which way'' mark is erased by means of a suitable measurement
---\emph{quantum erasure} \cite{scully82}---, interferences can be
somehow restored.

The acquisition of ``which way'' information in an interferometric device is
always accompanied by a reduction of the interference fringe visibility.
This well known, \emph{qualitative} observation has been recently translated into
modern and \emph{quantitative} expressions of Bohr's complementarity
\cite{GY88,englert,Bjo,Durr00}.
These quantitative complementarity expressions, as well as the
quantum eraser effect, have been investigated and confirmed in a variety of recent interferometric
experiments with atoms \cite{Durr00} or photon pairs \cite{qe-exp}.

The purpose of this contribution is to extend these
considerations, quantitative complementarity (or duality) and
quantum erasure, to neutral kaons. This is possible because a
clear analogy exists between the neutral kaon propagation in
free--space and the propagation of an interfering object in a
double--slit experiment. Indeed, a neutral kaon beam presents the
long known phenomenon of $K^0$--$\bar{K^0}$ oscillations, which
play the role of the interference fringes in a two--way
interferometer experiment. Similarly, the short-- and long--lived
kaon states, $K_S$ and $K_L$, being characterized by remarkably
different decay widths ($\Gamma_{S} >> \Gamma_{L}$) and distinct
propagation in free--space, are the analogs of the two separated
particle trajectories in interferometric devices. Therefore, the
better one can know if the propagation proceeds through the $K_S$
or $K_L$ component ---i.e., the more ``\emph{which width}''
information one can have---, the less pronounced will be the
$K_S$--$K_L$ interference and the visibility of the
$K^0$--$\bar{K^0}$ oscillations.

For single neutral kaons we deduce a generalized version of the quantitative
statement for complementarity proposed by Greenberger and Yasin
\cite{GY88} (Sec.~III.A).
Using instead entangled kaon pairs, another
quantitative duality statement by Englert \cite{englert}
and other authors \cite{Bjo,Durr00} is illustrated in Sec.~III.B.
Scully's quantum eraser \cite{scully82} for the neutral kaon system is
discussed in Sec.~IV.

\section{II.\, Neutral kaons}
\label{Intro}

Two alternative bases, each one associated to a possible
measurement, have to be considered for these neutral kaon analyses
\cite{bg1}. The strangeness basis, $\{K^0,\bar{K}^0\}$ with
$\langle K^0|\bar{K}^0\rangle =0$, is the appropriate one to
discuss strong production and reactions of kaons, as well as kaon
strangeness measurements. The latter are performed by inserting
along the kaon trajectory a piece of ordinary matter which induces
strangeness conserving kaon--nucleon strong interactions. The
neutral kaon is then destroyed after being identified either as a
$K^0$ or a $\bar{K}^0$ in a typically projective measurement.

The second basis consists of the $K_S$ and $K_L$ states having well
defined masses $m_{S(L)}$ and decay widths $\Gamma_{S(L)}$; it is
the appropriate one to discuss neutral kaon propagation in free--space,
with:
\begin{equation}
\label{evol}
|K_{S(L)}(\tau)\rangle = e^{-i\lambda_{S(L)}\tau} |K_{S(L)}\rangle
\end{equation}
and $\lambda_{S(L)}=m_{S(L)}-i\Gamma_{S(L)}/2$, as well as
lifetime measurements. The $K_S$ and $K_L$ eigenstates do not
oscillate into each other in time, but, since $\Gamma_S \simeq
579\, \Gamma_L$, the short--lived component of a given neutral
kaon extincts much faster than the long--lived one. Because of
this lifetime ``mark'', knowing if this kaon has propagated in
free--space \emph{either} as  $K_S$ \emph{or} $K_L$ is thus
possible by detecting at which time it decays. If kaons decaying
before $\simeq 4.8\, \tau_{S}$ after production are identified as
$K_S$'s and those surviving after this time as $K_L$'s,
misidentifications amount to only a few parts in $10^{-3}$
\cite{bg1}. 

The following relationship between the two kaon bases:
\begin{eqnarray}
\label{basis}
|K^0\rangle &=&\frac{1}{\sqrt{2}} \big[|K_S\rangle + |K_L\rangle \big] , \\
|\bar{K}^0\rangle &=&\frac{2}{\sqrt{2}} \big[|K_S\rangle - |K_L\rangle \big] , \nonumber
\end{eqnarray}
is valid when the small CP violating effects are neglected.
Note that this is a good approximation for our purposes
since these effects are of the same order as
the previously mentioned $K_S$ vs $K_L$  misidentifications.
Note also that the strangeness measurements and lifetime
observations completely exclude each other: the former require the
insertion of nucleonic matter, the latter propagation in free--space.
Bohr's complementarity principle is thus at work: if strangeness
(lifetime) is known, both  outcomes for lifetime (strangeness) are equally
probable.

\section{III.\, Quantitative complementarity}
\label{quant}
\subsection{A.\, Single kaons}
\label{quant-single}
Eqs.~(\ref{evol}) and (\ref{basis}) imply that
a kaon state which is
produced at time $\tau=0$ as a $K^0$ evolves in (proper) time according to the
expression:
\begin{equation}
\label{statet}
|K^0 (\tau)\rangle = \frac{1}{\sqrt{2}}
\big[e^{-i\lambda_S \tau} |K_S\rangle  + e^{-i\lambda_L \tau} |K_L\rangle \big] .
\end{equation}
By normalizing to kaons surviving up to time $\tau$, the previous state can be
more conveniently rewritten as:
\begin{eqnarray}
\label{statetn}
|K^0 (\tau)\rangle &=& \frac{|K_S\rangle  + e^{-{1\over 2} \Delta \Gamma \tau}
e^{-i \Delta m \tau} |K_L\rangle}
{\sqrt{1 + e^{-\Delta \Gamma \tau}}}  \nonumber \\
&\equiv &\frac{1}{\sqrt{2}} \big[|K_S(\tau)\rangle + |K_L(\tau)\rangle \big] \nonumber
\end{eqnarray}
where $\Delta \Gamma \equiv \Gamma_L-\Gamma_S < 0$ and $\Delta
m\equiv m_L-m_S$, or as:
\begin{eqnarray}
\label{statetnS}
|K^0 (\tau)\rangle &=&
{1 + e^{-{1\over 2}\Delta \Gamma \tau} e^{-i \Delta m \tau}
\over \sqrt{2\left(1 + e^{-\Delta \Gamma \tau}\right)}}\, |K^0\rangle \nonumber \\
&&+ {1 - e^{-{1\over 2}\Delta \Gamma \tau} e^{-i \Delta m \tau}
\over \sqrt{2\left(1 + e^{-\Delta \Gamma \tau}\right)}}\, |\bar{K^0}\rangle , \nonumber
\end{eqnarray}
in the $\{K^0,\bar{K^0}\}$ basis.

One thus easily obtains
the following $\tau$--dependent transition probabilities:
\begin{eqnarray}
\label{s1}
\big|\langle K^0|K^0(\tau)\rangle\big|^2 &=&\frac{1}{2}
\big\{1+{\cal V}_0(\tau)\, {\rm cos}(\Delta m\, \tau)\big\} ,  \\
\label{s2}
\big|\langle \bar{K}^0|K^0(\tau)\rangle\big|^2 &=& \frac{1}{2}
\big\{1-{\cal V}_0(\tau)\, {\rm cos}(\Delta m\, \tau)\big\} ,
\end{eqnarray}
where:
\begin{equation}
\label{visib}
{\cal V}_0(\tau)=\frac{1}{\displaystyle \cosh\left(\frac{\Delta\Gamma\, \tau}{2}\right)}
\end{equation}
is the time--dependent fringe visibility of the $\bar{K}^0$--$K^0$ oscillations.
On the contrary, no $K_S$--$K_L$ oscillations are expected:
\begin{eqnarray}
\label{l1}
\big|\langle K_L|K^0(\tau)\rangle\big|^2 &=& \frac{1}{2}\, \big|\langle
K_L|K_{L}(\tau)\rangle\big|^2 \\
&=& \frac{1}{1+e^{\Delta \Gamma \, \tau}} , \nonumber \\
\label{l2}
\big|\langle K_S|K^0(\tau)\rangle\big|^2 &=& \frac{1}{2}\, \big|\langle
K_S|K_{S}(\tau)\rangle\big|^2 \\
&=& \frac{1}{1+e^{-\Delta \Gamma \, \tau}} . \nonumber
\end{eqnarray}

These observations admit the following interpretation, as
discussed in \cite{ours2,ours1}. As soon as a  $K^0$ is produced,
it starts propagating in free--space in the coherent superposition
of $K_S$ and $K_L$ [see Eq.~(\ref{statet})] and mimics the
two--way propagation of any quantum system beyond a symmetrical
double--slit. In the familiar double--slit case, the system
follows the two paths without ``jumping'' from one to the other in
the same way as $K_S$--$K_L$ oscillations are ``forbidden''. In
the kaon case, however, there are not two separated trajectories
but a single path comprising \emph{automatically} (i.e., with no
need of any double--slit like apparatus) the two components $K_S$
and $K_L$. As previously stated, these two interfering components
are marked by their different decay widths. But this is an
intrinsic mark on each component which is {\it automatically}
given by Nature.

For initial $K^0$'s surviving up to time $\tau$, the probabilities for $K_S$ and $K_L$
propagation are
\begin{eqnarray}
w_S(\tau)&\equiv& \big|\langle K_S|K^0(\tau)\rangle\big|^2
=\frac{1}{1+e^{-\Delta \Gamma \, \tau}} , \nonumber \\
w_L(\tau)&\equiv& \big|\langle K_L|K^0(\tau)\rangle\big|^2
=\frac{1}{1+e^{\Delta \Gamma \, \tau}} . \nonumber
\end{eqnarray}
From these expressions one obtains the ``width predictability''
(corresponding to the ``path predictability'' defined in Ref.\cite{GY88}):
\begin{equation}
\label{predic}
{\cal P}(\tau)\equiv \big|w_S(\tau)-w_L(\tau)\big|=
\left|\tanh\left(\frac{\Delta \Gamma \tau}{2}\right)\right| .
\end{equation}
${\cal P}(\tau)$ quantifies the a priori (i.e., before any
measurement is performed) ``which width'' knowledge we have
from the fact that a kaon
which was created as a $K^0$ at time $\tau=0$ has survived up to time $\tau$.
At $\tau =0$,
both components start propagating with the same probability, $w_S(0)=w_L(0)=1/2$, and the
``width predictability'' vanishes, ${\cal P}(0)=0$. In other words,
there is no information on which component actually propagates and
the visibility of strangeness oscillations is maximal, ${\cal V}_0(0) =1$.
However, since the $K_S$ and $K_L$ components are intrinsically
``marked'' by their different lifetimes, ``which width'' information is obtained for
initial $K^0$'s  surviving up to time $\tau$ and the corresponding oscillation visibility
is thus reduced.
The fringe visibility and ``width predictability'' of Eqs.~(\ref{visib}) and (\ref{predic})
fulfill the equation:
\begin{equation}
\label{duality-gy}
{\cal V}^2_0(\tau) + {\cal P}^2(\tau) =1 ,
\end{equation}
which can be viewed as a
generalization of the quantitative duality relation of Ref.\cite{GY88}
to situations where ${\cal P}(\tau)$ and ${\cal V}_{0}(\tau)$ are not
constants.

Note that the same expressions of Eqs.~(\ref{visib}) and (\ref{predic})
for ${\cal V}_0(\tau)$ and ${\cal P}(\tau)$
are valid when the kaon state produced at time $\tau=0$ is a $\bar{K}^0$.
In this case, as well as in the previous one starting with an initial
$K^0$, the state remains quantum mechanically pure.
The two terms in Eq.~(\ref{duality-gy}) add up to one rather than
fulfilling the less stringent relation ${\cal V}^2_0 + {\cal P}^2 < 1$
for mixed states.

Recent measurements by the CPLEAR Collaboration \cite{CPLEARreview} have been
interpreted \cite{ours1} in terms of the quantitative duality we have just discussed.
The proton--antiproton annihilation processes,
$p \bar p \to K^- \pi^+ K^0$ or $p \bar p \to K^+ \pi^- \bar{K^0}$, were used
to produce either a $K^0$ or a $\bar{K}^0$ initial state,
which were then allowed to propagate in free--space.
In a first experiment \cite{CPLEARlepton},
strangeness oscillations were observed by detecting semileptonic
neutral kaon decays. In another experiment
\cite{CPLEARstrong}, $K^0$--$\bar{K}^0$ oscillations
were observed via strangeness measurements monitored
by kaon--nucleon strong interactions.
The experimental data collected in both
CPLEAR experiments are seen to fulfil the statement for quantitative
duality of Eq.~(\ref{duality-gy}).
Further experiments at the operating $\phi$--factory Daphne \cite{handbook},
which copiously produces neutral kaons via strong $\phi \to K^0 \bar{K}^0$
decays, could provide interesting and more accurate tests.

\subsection{B.\, Entangled kaon pairs}
\label{quant-entang}

The a priori knowledge on the kaon lifetimes and thus its time
evolution, expressed in the previous section in terms of the
predictability ${\cal P}(\tau)$, comes exclusively from knowing
that the kaon remains undecayed at time $\tau$. This knowledge can
be obviously improved if a measurement is performed on the kaon
state \cite{englert,Bjo,Durr00}. However, since strangeness
measurements and lifetime observations are completely destructive
and no other projective measurement is possible on neutral kaons,
two--particle entanglement is necessary to achieve such a purpose.
Working with entangled kaons, one can perform a measurement on one
member (the \emph{meter} kaon) which allows one to increase the
information on the propagation mode of its partner (the
\emph{object} kaon) without this being annihilated.

To this aim, consider the decay of the $\phi$--meson \cite{handbook}
(or, alternatively, $S$--wave $p\bar p$ annihilation \cite{CPLEAR}) into
$K^0 \bar{K}^0$ pairs. Just after the decay, i.e., at time $\tau=0$,
one has the following maximally entangled state:
\begin{eqnarray}
\label{entangled}
|\phi(0)\rangle & = & \frac{1}{\sqrt 2}\big[
|K^0\rangle_l |\bar{K}^0\rangle_r - |\bar{K}^0\rangle_l |K^0\rangle_r\big] \nonumber \\
& = & \frac{1}{\sqrt 2}\big[
|K_L\rangle_l |K_S\rangle_r - |K_S\rangle_l |K_L\rangle_r\big] , \nonumber
\end{eqnarray}
where $l$ and $r$ denote the ``left'' and ``right'' kaon directions of motion.
In the lifetime basis, the state evolution up to
time $\tau_l$ ($\tau_r$) along the left (right) beam is given by:
\begin{eqnarray}
\label{notnorm}
&&|\phi(\tau_l,\tau_r)\rangle = \frac{1}{\sqrt 2}
\biggl\{
e^{-i(\lambda_L\tau_l+\lambda_S\tau_r)}|K_L\rangle_l|K_S\rangle_r \nonumber \\
&&\hphantom{|\phi(\Delta\tau)\rangle = N(\tau_l)a}
 -e^{-i(\lambda_S\tau_l+ \lambda_L\tau_r)}|K_S\rangle_l|K_L\rangle_r\biggl\} . \nonumber
\end{eqnarray}

It is convenient to consider only kaon pairs with both left and right members surviving up to
$\tau_{l}$ and $\tau_{r}$. These are described by the normalized state:
\begin{eqnarray}
\label{time}
|\phi(\Delta\tau)\rangle&=& \frac{1}{\sqrt {1+e^{\Delta\Gamma \Delta\tau}}}
\biggl\{|K_L\rangle_l|K_S\rangle_r \\
&& \qquad -e^{i \Delta m \Delta\tau} e^{{1 \over 2} \Delta \Gamma \Delta\tau}
|K_S\rangle_l|K_L\rangle_r\biggl\} ,\nonumber
\end{eqnarray}
where $\Delta\tau\equiv \tau_l-\tau_r$. Eq.~(\ref{time}) can be rewritten as:
\begin{eqnarray}
\label{timestrangeness}
&&|\phi(\Delta\tau)\rangle =
\frac{1}{\sqrt {2(1+e^{\Delta\Gamma \Delta\tau})}} \\
&&\times \biggl\{|K^0 \rangle_l|K_S\rangle_r -|\bar K^0 \rangle_l|K_S\rangle_r \nonumber \\
&& -e^{i \Delta m \Delta\tau} e^{{1 \over 2} \Delta \Gamma \Delta\tau}
\left[|K^0 \rangle_l|K_L\rangle_r + |\bar K^0
\rangle_l|K_L\rangle_r\right]\biggl\} , \nonumber
\end{eqnarray}
or equivalently as:
\begin{eqnarray}
\label{bothstrangeness}
|\phi(\Delta\tau)\rangle &=& \frac{1}{2 \sqrt {1+e^{\Delta\Gamma \Delta\tau}}} \\
&& \times \biggl\{\left(1-e^{i \Delta m \Delta\tau} e^{{1 \over 2} \Delta \Gamma
\Delta\tau}\right) \nonumber \\
&&\times \big[|K^0\rangle_l|K^0\rangle_r-|\bar K^0\rangle_l|\bar K^0\rangle_r\big]  \nonumber \\
&& +\left(1+e^{i \Delta m \Delta\tau} e^{{1 \over 2} \Delta \Gamma
\Delta \tau}\right) \nonumber \\
&&\times \big[|K^0\rangle_l|\bar K^0\rangle_r-|\bar
K^0\rangle_l|K^0\rangle_r\big] \biggl\} .\nonumber
\end{eqnarray}
Eqs.~(\ref{time})--(\ref{bothstrangeness})
immediately supply the various joint
probabilities $P(K_{l},K_{r})$ for detecting a $K_{l}$ ($K_{r}$) on the left
(right) at time $\tau_{l}$ ($\tau_{r}$)
with $K_{l,r} = K^0, \bar K^0 , K_{S}$ or $K_{L}$.

Let us now consider the following two experimental arrangements
\cite{ours2}. In both arrangements one measures the strangeness of
the left moving kaon  by inserting a dense slab of (nucleonic)
matter at different distances along its trajectory; since each
distance corresponds to a given time--of--flight $\tau_l$, one can
measure at different  values of $\tau_l$ and look for strangeness
oscillations of this (object) kaon. On the other hand, the
measurement on the right moving (meter) kaon is always performed
at a \emph{fixed} time $\tau^0_r$; but on this kaon, one of two
alternative measurements, $\cal M =\cal S$ or $\cal L$, is
performed by either inserting or not a strangeness detector at
$\tau^0_r$. In the first case, $\cal M =\cal S$, strangeness is
measured on both kaons and four $\tau_l$--dependent joint
probabilities $P(K_{l},K_{r})$ with $K_{l,r} = K^0$ or $\bar K^0$
can be recorded.
In the second case, by removing the strangeness detector one
observes the lifetime of the right moving kaon, $\cal M =\cal L$,
and measures the four joint probabilities $P(K_{l},K_{r})$ with
$K_{l} = K^0$ or $\bar K^0$ and  $K_{r} = K_{S}$ or $K_{L}$ (for
explicit expressions and details see Sec.IV.~).

Following Englert \cite{englert},
we can then define the ``which width knowledge'' for the object kaon,
${\cal K_M}(\tau_l)$,
which depends on the measurement (either $\cal M =\cal S$  or $\cal M =\cal L$)
performed on the meter kaon. Note however that an experiment can never decrease
the information provided by the
formerly discussed predictability ${\cal P}(\tau_l)$,  and one necessarily has
${\cal K_M}(\tau_l) \geq {\cal P}(\tau_l)
\equiv |\tanh (\Delta \Gamma\, \tau_l/2)|$.
One can similarly consider the corresponding visibility of the object kaon
strangeness oscillations, ${\cal V_M}(\tau_l)$, which now satisfies
${\cal V_M}(\tau_l) \leq {\cal V}_0^2(\tau_l)\equiv 1/\cosh(\Delta \Gamma\,
\tau_l/2)$. It will be seen that ${\cal K_M}(\tau_l)$ and ${\cal V_M}(\tau_l)$ are
linked by a new quantitative duality relation
\cite{Bjo,Durr00}:
\begin{equation}
\label{duality-dr}
{\cal V^{\rm 2}_M}(\tau_l)+{\cal K^{\rm 2}_M}(\tau_l)=1 .
\end{equation}

Without the strangeness detector on the right beam, $\cal M =\cal L$,
one can observe the decay of the freely propagating meter kaon, which
will be identified either as $K_{S}$ or $K_{L}$.
The acquisition of this ``which width'' information implies the
corresponding one for the object kaon and therefore strangeness
oscillations (in $\tau_l$)
should not be visible:
\begin{equation}
{\cal V_L}(\tau_l)=0 \, \, \, \forall \, \tau_l ,
\end{equation}
for any of the four possible joint detection probabilities
$P(K_{l},K_{r})$ with $K_{l} = K^0$ or $\bar K^0$ and  $K_{r} = K_{S}$
or $K_{L}$. This is immediately seen from Eq.~(\ref{timestrangeness}).
Using Eq.~(\ref{time}) one obtains the maximum value for the object kaon
``which width knowledge'' \cite{englert}:
\begin{eqnarray}
{\cal K_L}(\tau_l)\equiv \left|P(K_S(\tau_l),K_S(\tau^0_r))-
P(K_L(\tau_l),K_S(\tau^0_r)) \right| &&\nonumber \\
+\left|P(K_S(\tau_l),K_L(\tau^0_r))- P(K_L(\tau_l),K_L(\tau^0_r))
\right| =1 \, \, \forall \, \tau_l , && \nonumber
\end{eqnarray}
where now the joint probabilities explicitly show their dependence
on left and right mesurement times $\tau_{l}$ and $\tau_{r}^0$.
The quantitative duality relation of Eq.~(\ref{duality-dr}) is
then trivially satisfied.

Considering instead a strangeness measurement along the right beam,  $\cal M =\cal S$,
one obtains the
four joint probabilities $P(K_{l},K_{r})$ with $K_{l,r} = K^0$ or $\bar K^0$
which show the $\tau_l$--dependent
strangeness oscillations and anti--oscillations immediately
deducible from Eq.~(\ref{bothstrangeness}),
with visibility:
\begin{equation}
\label{vis-s}
{\cal V_S}(\tau_l)=\frac{1}{\displaystyle \cosh\left(\frac{\Delta\Gamma\,
(\tau_l-\tau^0_r)}{2}\right)} .
\end{equation}
This visibility is maximal
(${\cal V_S}=1$) for $\tau_l=\tau^0_r$ (no ``which width'' information
available), but the strangeness oscillations disappear (${\cal V_S}\to 0$)
for $\tau_l\to \infty$ (full ``which width'' information).
The ``which width'' information on the object kaon, which is available as a result of
the joint strangeness measurement, can be expressed in terms of the
``which width knowledge'' \cite{englert}:
\begin{eqnarray}
{\cal K_S}(\tau_l)&\equiv& \left|P[K_S(\tau_l),K^0(\tau^0_r)]-P[K_L(\tau_l),K^0(\tau^0_r)]\right|
\nonumber \\
&+&\left|P[K_S(\tau_l),\bar{K}^0(\tau^0_r)]-P[K_L(\tau_l),\bar{K}^0(\tau^0_r)]\right| \nonumber \\
&=&\left|\tanh\left(\frac{\Delta \Gamma (\tau_l-\tau^0_r)}{2}\right)\right| , \nonumber
\end{eqnarray}
which, together with the visibility (\ref{vis-s}),
satisfies again the quantitative duality requirement (\ref{duality-dr}).

Note that the two types of alternative measurements performed on the
right moving kaon are clearly different. If
strangeness is measured, $\cal M =\cal S$, no ``which width'' information
is obtained in addition to the information that was already known a priori, thus ${\cal
K_S}(\tau_l) \equiv {\cal P}(\tau_l)$. In the other case,  $\cal M =\cal
L$, the opposite situation is achieved and the ``which width'' knowledge
is maximally increased, ${\cal
K_L}(\tau_l) \equiv 1 > {\cal P}(\tau_l)$. According to the definition
of Ref.~\cite{englert}, this corresponds to a ``width
distinguishability'' ${\cal D}(\tau_l) \equiv 1$.

\section{IV.\, Quantum eraser}
\label{Eraser}
The discussion of Sec.III.~can be reinterpreted
in terms of Scully's \emph{quantum eraser} \cite{scully82}, which
consists of three different steps \cite{kwiat2}.

In the first step one has to start with a coherent kaon
superposition of $K_S$ and $K_L$ states such as the one in
Eq.~(\ref{statet}), which provides $K_S$--$K_L$ interference
effects showing the $K^0$--$\bar{K}^0$ oscillations of
Eqs.~(\ref{s1}) and (\ref{s2}). These oscillations become less
pronounced with time as a consequence of the increase of the $K_S$
and $K_L$ ``path predictability''.

In the second step one considers entangled kaon pairs in the state
(\ref{time}). If the right going meter kaon is free to propagate
in space, its lifetime ``mark'' is operative and precludes the
observation of any strangeness oscillation of the entangled
partner, the object kaon. The following joint probabilities are
then measured:
\begin{eqnarray}
\label{no-int}
P[K^0(\tau_l),K_S(\tau^0_r)]&=&P[\bar{K}^0(\tau_l),K_S(\tau^0_r)] \\
&=&\frac{1}{2(1+e^{\Delta\Gamma (\tau_l-\tau^0_r)})} , \nonumber \\
P[K^0(\tau_l),K_L(\tau^0_r)]&=&P[\bar{K}^0(\tau_l),K_L(\tau^0_r)] \nonumber \\
&=&\frac{1}{2(1+e^{-\Delta\Gamma (\tau_l-\tau^0_r)})} . \nonumber
\end{eqnarray}

The possibility to obtain full ``which width'' information for the object kaon can be
precluded by the third step of the quantum eraser. In order to erase the
lifetime ``mark'' of the meter kaon, it has to be measured
in the $\{K^0,\bar{K}^0\}$ basis of Eqs.~(\ref{basis}); strangeness oscillations
and their complementary anti--oscillations appear when sorting the
jointly detected events according to the following probabilities:
\begin{eqnarray}
\label{int}
P[K^0(\tau_l),K^0(\tau^0_r)]&=&P[\bar{K}^0(\tau_l),\bar{K}^0(\tau^0_r)] \\
&=&\frac{1-{\cal V}(\tau_l)\cos [\Delta m(\tau_l-\tau^0_r)]}{4} , \nonumber \\
P[K^0(\tau_l),\bar{K}^0(\tau^0_r)]&=&P[\bar{K}^0(\tau_l),K^0(\tau^0_r)] \nonumber \\
&=&\frac{1+{\cal V}(\tau_l)\cos [\Delta m(\tau_l-\tau^0_r)]}{4}. \nonumber
\end{eqnarray}
Note that this third step restores
the same $K_S$--$K_L$ interference effects [see Eqs.(\ref{s1}) and (\ref{s2})]
of the first step, where single kaon states [see Eq.~(\ref{statet})] are used.

Note also that the joint probabilities of Eqs.~(\ref{no-int})
and (\ref{int}), ${\cal K_M}$ and ${\cal V_M}$
are all of them even functions of $\tau_l-\tau^0_r$.
The  erasure of the lifetime ``mark'' on the meter kaon can thus be delayed to times after the
object kaon has been detected ($\tau^0_r>\tau_l$). This ``delayed choice'' mode
captures the essential feature of the quantum eraser \cite{scully82},
which is a proper sorting of the jointly detected events.


\section{V.\, Conclusions}

To summarize, we have discussed new quantitative formulations of Bohr's complementarity
for free--space propagation of single and entangled pairs of neutral kaons.
Single neutral kaons allow for a generalization and clear application of the Greenberger and
Yasin duality relation [Eq.~(\ref{duality-gy})]. In this case,
the two recent CPLEAR experiments of Refs.~\cite{CPLEARlepton,CPLEARstrong}
admit a transparent interpretation in terms of
the quantitative complementarity requirement of Eq.~(\ref{duality-gy}) and the data
fully agree with this equality.
Entangled $K^0\bar{K}^0$ pairs allows for the more interesting
complementarity test in terms of ``width knowledge'' and ``width distinguishability''
suggested by Englert and other authors.
Experimental tests on this issue could be performed at the
$\phi$--factory Daphne \cite{handbook}.

The neutral kaon system reveals also suitable for an
optimal demonstration \cite{scully82,kwiat2} of quantum erasure: (1)
the ``which width'' information is carried by a system (the meter kaon)
distinct and spatially separated from the
interfering system (the object kaon); (2) as a consequence, the erasure operation can
be performed in the ``delayed choice'' mode ($\tau_l<\tau^0_r$);
(3) single--particle states (as opposed to coherent states) are
detected on each side;
(4) in spite of the need to entangle the object kaon with another system
(the meter kaon), quantum erasure allows one to restore
the same $K_S$--$K_L$ interference phenomenon exhibited by a single kaon state
produced as $K^0$ or $\bar{K}^0$.

An experimental test of the marking and erasure operations
we have discussed is desirable and should be feasible at $\phi$--factories and $p\bar{p}$
machines. Actually, the CPLEAR collaboration \cite{CPLEAR}
has already done part of the
work required: the two experimental points (for $|\tau_l-\tau^0_r|$= 0 and
$1.2\, \tau_S$)
collected by this experiment are compatible with the
joint strangeness oscillations predicted by Eqs.~(\ref{int}).
New measurements confirming with better precision these
oscillations for a larger range of $\tau_l-\tau^0_r$ values,
as well as the non--oscillating behaviour when ``which width'' information
is in principle available [see Eq.~(\ref{no-int})],
are needed for a full demonstration of quantitative duality and
the quantum eraser with neutral kaons.

\section{Acknowledgements}
This work has been supported by  EURIDICE HPRN--CT--2002--00311,
BFM--2002--02588, Austrian Science Foundation (FWF) SFB 015P06 and
INFN.


\end{document}